\documentclass[12pt]{article}
\usepackage{natbib}
\usepackage{graphicx}
\usepackage{times}
\usepackage{geometry}
\usepackage[utf8]{inputenc}
\usepackage{enumitem,amssymb}
\usepackage{ragged2e}
\newlist{thematic}{itemize}{8}
\setlist[thematic]{label=$\square$}
\usepackage{pifont}
\newcommand{\xmark}{\ding{55}}%

\usepackage{titlesec}
\titlespacing*{\section}{0pt}{1.5ex}{1.2ex}

\def\fun#1#2{\lower3.6pt\vbox{\baselineskip0pt\lineskip.9pt
        \ialign{$\mathsurround=0pt#1\hfill##\hfil$\crcr#2\crcr\sim\crcr}}}

\begin{document}
{\raggedright
\huge
Astro2020 Science White Paper \linebreak

Observing Galaxy Evolution in the Context of Large-Scale Structure \linebreak
\normalsize

\vspace{-1ex}
\noindent \textbf{Thematic Areas:} \hspace*{60pt} $\square$ Planetary Systems \hspace*{10pt} $\square$ Star and Planet Formation \hspace*{20pt}\linebreak
$\square$ Formation and Evolution of Compact Objects \hspace*{31pt} $\square$ Cosmology and Fundamental Physics \linebreak
  $\square$  Stars and Stellar Evolution \hspace*{1pt} $\square$ Resolved Stellar Populations and their Environments \hspace*{40pt} \linebreak
  \xmark    Galaxy Evolution   \hspace*{45pt} $\square$             Multi-Messenger Astronomy and Astrophysics \hspace*{65pt} \linebreak

\vspace{-1ex}
\textbf{Principal Author:}

Name:	Mark Dickinson
 \linebreak						
Institution:  NOAO
 \linebreak
Email:  med@noao.edu
 \linebreak
Phone:  520-318-8531
 \linebreak

\vspace{-1ex}
\textbf{Co-authors:} 
\linebreak
Yun Wang (Caltech/IPAC), James Bartlett (NASA JPL), Peter Behroozi (Arizona), Jarle Brinchmann (Leiden Observatory; Porto), 
Peter Capak (Caltech/IPAC),
Ranga Chary (Caltech/IPAC), Andrea Cimatti (Bologna; INAF), 
Alison Coil (UC San Diego),
Charlie Conroy (CfA), Emanuele Daddi (CEA, Saclay), Megan Donahue (Michigan State University), Peter Eisenhardt (NASA JPL), Henry C.\ Ferguson (STScI), 
Steve Furlanetto (UCLA),
Karl Glazebrook (Swinburne),
Anthony Gonzalez (Florida),
George Helou (Caltech/IPAC), 
Philip F.\ Hopkins (Caltech),
Jeyhan Kartaltepe (RIT),
Janice Lee (Caltech/IPAC),
Sangeeta Malhotra (NASA GSFC), 
Jennifer Marshall (Texas A\&M),
Jeffrey A.\ Newman (Pittsburgh), Alvaro Orsi (CEFCA), James Rhoads (NASA GSFC), Jason Rhodes (NASA JPL), Alice Shapley (UCLA), Risa H.\ Wechsler (Stanford/KIPAC; SLAC)
\linebreak

\vspace{-1ex}
\textbf{Abstract:}
\linebreak
Galaxies form and evolve in the context of their local and large-scale environments.   Their baryonic content that we observe with imaging and spectroscopy is intimately connected to the properties of their  dark matter halos, and to their location in the ``cosmic web'' of large-scale structure.   Very large spectroscopic surveys of the local universe (e.g., SDSS and GAMA) measure  galaxy positions (location within large-scale structure), statistical clustering (a direct constraint on dark matter halo masses), and spectral features (measuring physical conditions of the gas and stars within the galaxies, as well as internal velocities).   Deep  surveys with the James Webb Space Telescope (JWST) will revolutionize spectroscopic measurements of redshifts and spectral properties for galaxies out to the epoch of reionization, but with numerical statistics and over cosmic volumes that are too small to map large-scale structure and to constrain halo properties via clustering.  Here, we consider advances in understanding galaxy evolution that would be enabled by very large  spectroscopic surveys at high redshifts:   very large numbers of galaxies (outstanding statistics) over  large co-moving volumes (large-scale structure on all scales) over broad redshift ranges (evolution over most of cosmic history). The required observational facility can be established as part of the probe portfolio by NASA within the next decade.
}
\pagebreak

% \addtolength{\parskip}{1 ex}
% \setlength{\parindent}{2 ex}

\section{Introduction}

The initial conditions for galaxy formation come from the quantum fluctuations that were imprinted on the dark matter distribution during inflation. The large-scale distribution of galaxies results from the gravitational evolution and hierarchical clustering of these fluctuations. In today's era of precision cosmology, we believe that we understand the growth of structure in a universe of cold dark matter and dark energy, and can map this over cosmic time with sophisticated numerical simulations. However, the galaxies that we see are not simply dark matter halos. Baryonic physics makes them far more complex, and we are still far from understanding how galaxies form and develop in the context of an evolving ``cosmic web" of dark matter, gas and stars. Gas flows along intergalactic filaments defined by the skeleton of the dark matter distribution. It cools and condenses into dark matter halos, forming stars that produce heavy elements that further alter the evolutionary history of the baryons.  A full understanding of galaxy evolution will not emerge until models are constrained by rich observational data over a broad swath of cosmic history, during which galaxies and cosmic structure were growing together.

Spectroscopy for very large galaxy samples over large cosmic volumes is critical for achieving this full understanding.  In the local ($z \sim 0$) universe, this has been amply demonstrated by very large, rich spectroscopic surveys like SDSS and GAMA.  Spectra provide diagnostics of important physical properties of gas and stars within galaxies, while also precisely locating those galaxies within their environmental context.  Large surveys enable robust statistical analyses that span a wide range of galaxy properties (type, mass, morphology, star formation rate, chemical abundance, etc.) and environment (from voids to rich clusters).   At higher redshifts ($z \sim 0.7$), surveys like zCOSMOS \citep{Lilly09}, PRIMUS \citep{Coil11} and VIPERS \citep{Guzzo14} have begun to explore galaxies in the context of large-scale structure.   At still higher redshifts, deep  ``pencil-beam''  spectroscopy on 8-10m telescopes measures redshifts out to $z \approx 8$, but with sample sizes and survey volumes that are orders of magnitude too small to robustly connect galaxy evolution to environment.  DESI, PFS and MOONS will measure large samples over large volumes, but these are (primarily) optical instruments:  the number of galaxies measured will drop steeply at $z \gg 1$, and these instruments will make only limited measurements of the optical rest-frame nebular emission lines that are critical for measuring ISM conditions or kinematics, let alone continuum and absorption lines to diagnose stellar population properties.   Slitless near-infrared spectroscopy from Euclid and WFIRST can survey large volumes, but with limited sensitivity and wavelength coverage and very low spectral resolution, leading to a very blurry view of the 3D galaxy distribution over restricted ranges of redshift (Fig.~1). JWST NIRSpec will revolutionize deep field spectroscopy with full wavelength coverage from 1 to 5 $\mu$m, measuring optical rest frame emission lines out to the epoch of reionization, but only over very narrow fields and in samples of hundreds or thousands, not millions.    

Here, we consider the advances in our understanding of galaxy evolution that could be achieved with a ``time lapse SDSS'' that obtains spectra for vast numbers of galaxies over very large volumes spanning most of cosmic history.

\section{Galaxy Evolution and the Cosmic Web}

Galaxy properties correlate with those of the underlying dark matter halos: their masses,  spins, and  positions within the cosmic web.  The challenge of deriving dark matter halo masses for high redshift galaxies is best met with very large spectroscopic surveys that can measure galaxy clustering for galaxies binned by many other observable/inferable parameters, such as stellar mass, star formation rate, or morphology (Fig.~2).  The \textit{stellar mass---halo mass relationship} (SMHMR) \citep[e.g.,][]{Moster13,Behroozi13} measures the efficiency with which galaxies turn incoming gas into stars, and is a key probe of the strength of feedback from stars and supermassive black holes.  Today we have only rough estimates of the SMHMR from abundance matching, clustering, and weak lensing (\citealt{Behroozi13}, \citealt{Kravtsov13}, \citealt{Hudson15}). Very large spectroscopic surveys can  measure redshift evolution of the SMHMR for massive halos (bias $b\gtrsim1$, Fig.~3a) as a function of galaxy properties, which is impossible to do via abundance matching. It is also clear that the SMHMR does not capture the entire influence of dark matter on galaxy properties.  For example, galaxy star formation rates correlate strongly with halo growth rates, a result which requires spectroscopic environmental measures to constrain \citep{Behroozi18}.

Additional constraints can be provided by \textit{group catalogs}, wherein halo masses are measured for individual galaxies based on spectroscopically-identified satellite galaxy counts (Fig.~3b).  These catalogs will also provide information about whether a given galaxy is a satellite or not; this is important for interpreting whether galaxy properties arise mainly from internal processes or instead from interaction with a larger neighbor. 

Other open questions in galaxy formation concern how halo assembly affects galaxy properties and supermassive black hole activity.  Halo assembly is not directly measurable, but many assembly properties (e.g., concentration, spin, mass accretion rate) correlate strongly with \textit{environmental density} \citep[e.g.,][]{Lee17}.  Large spectroscopic surveys to $z \sim 3$ or beyond will be able to measure environmental densities (Fig.~3c). They can also measure average dark matter accretion rates for galaxies via the detection of the splash-back radius (a.k.a., turnaround radius) of their satellites as in \cite{More16} out to $z\sim5$ (Fig.~3d).

\begin{figure}
\centering
\vspace{-2ex}
\includegraphics[width=1.1\columnwidth,clip]{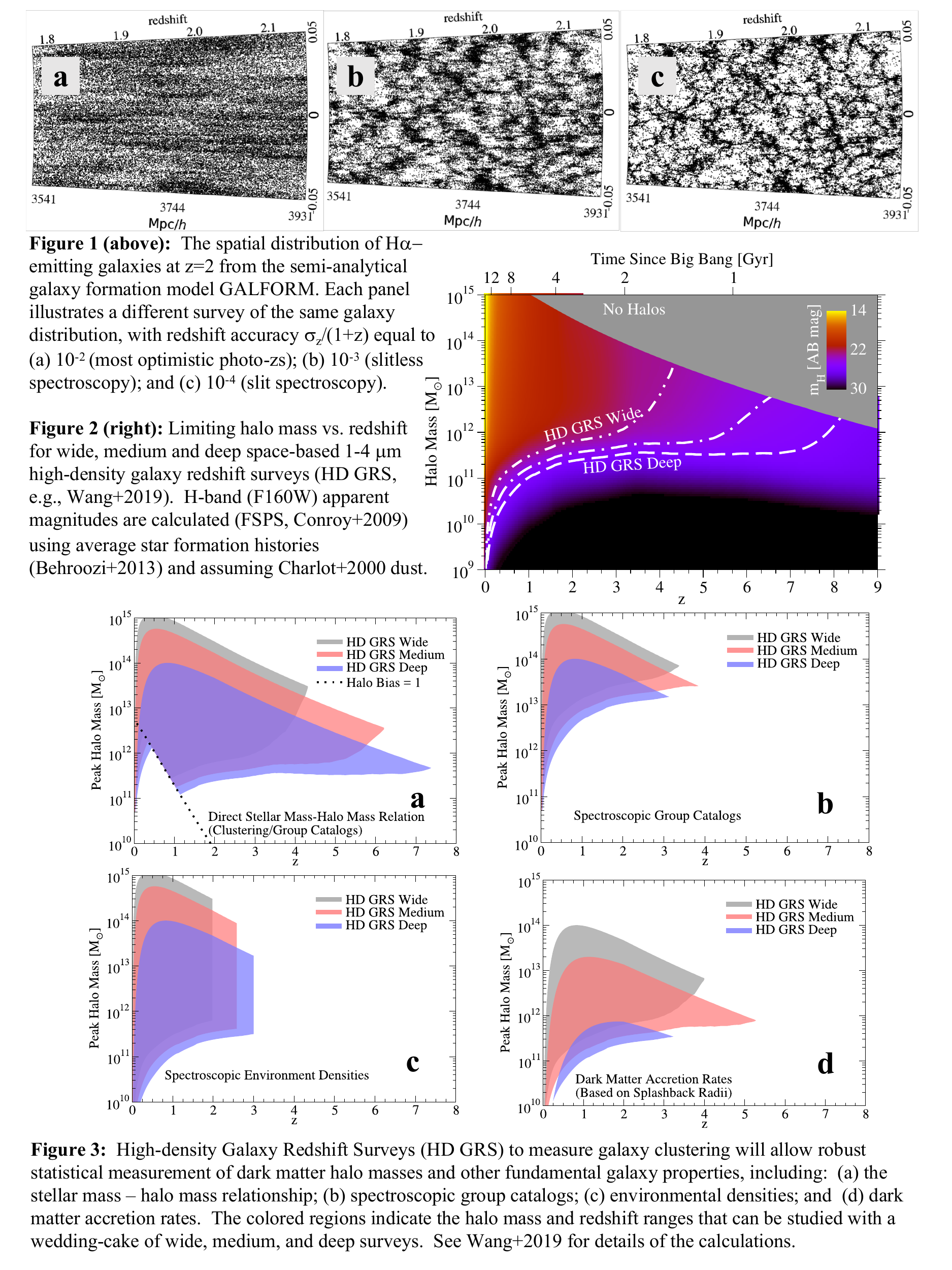}
\end{figure}

\section{Black Holes, AGN, and Feedback}

 The shape of the SMHMR  implies a characteristic mass at which galaxies most effectively convert gas into stars. At lower and higher masses, ``feedback" is invoked to prevent gas from cooling onto galaxies, or to expel gas, thus suppressing star formation efficiency. The most massive galaxies today are predominantly quiescent, with little active star formation, and with dominant bulges that universally contain supermassive black holes. These, when fueled, can power active nuclei which may dump tremendous energy into their environments. AGN feedback, perhaps coupled with environmental effects, may be the dominant process regulating star formation and growth for massive galaxies. Like cosmic star formation, the luminous output of AGN and (by inference) the black hole growth rate, peaked at high redshift.  Spectroscopy at 1-4 $\mu$m will be uniquely suited to identifying vast samples of AGN over a wide range of luminosity and redshift, using standard nebular excitation diagnostics (e.g., \citealt{Baldwin81}) and through detection of high-ionization emission lines.  This will connect AGN activity to local and large-scale environment with exquisite statistical accuracy that is only possible today in the local universe, and [OIII] luminosities will provide a critical measure of the distribution of accretion luminosities and black hole growth rates. 

\section{Early Galaxy Clusters and Proto-clusters}

The first  galaxy clusters, hosted within dark matter halos with masses of a few $\times 10^{13}$ $M_\odot$, are expected assemble at $z \sim$ 2-3, with space densities ranging from 10$^{-5}$ to $10^{-6}\,$Mpc$^{-3}$, corresponding to several per square degree. 
Observations of young clusters at these redshifts (\citealt{Gobat13};  \citealt{WangT16}) show that they host spectacular activity: from strong star formation to quenching, from AGN feedback to the morphological transformation of galaxies and the formation of ellipticals. AGN activity is likely to be prominent, following in parallel the rise of SFR and gas content in the Universe. In these conditions, interactions between the ICM and the cluster's environment through the inflow of pristine cold gas, the outflows originated from stellar and AGN's winds, and the deposition of warm plasma will shape the baryon distribution, the metal content and the energy budget of the structures (e.g., \citealt{Valentino15,Valentino16}).

A deep and densely-sampled infrared spectroscopic survey covering 100 square degrees to line flux limits of a few 10$^{-18}$ erg/s/cm$^2$) would confirm galaxy membership  within several hundred early galaxy groups and clusters at $2 < z < 3$, measuring galaxy spectral properties as well as cluster dynamics. At still higher redshifts, deep spectroscopy at 1 to 4$\mu$m can identify early groups ($M_\mathrm{halo} \sim 10^{13} M_\odot$) as yet still-collapsing proto-clusters at $z < 5$ and $< 7$ with H$\alpha$ and [OIII], respectively.   Groups of $10^{13}\, M_\odot$ have an expected density of $\sim 0.25$~deg$^{-2}$ at $5<z<7$.  Detailed characterization down the stellar mass function will require deep exposures to reach line flux limits of order a few $10^{-19}$ erg/s/cm$^2$. Observing such objects will be key to explore the early phases of environmental effects on galaxy formation and evolution as well as the beginning of structure formation (see e.g., \citealt{Orsi16}, and \citealt{Izquierdo17}). 

\section{Galaxy Kinematics}

Kinematic measurements for vast samples of high redshift galaxies would be invaluable for constructing a detailed picture of galaxy evolution in the context of large-scale structure.   Emission-line widths can be interpreted in concert with WFIRST imaging and structural properties to infer dynamical masses \citep{Price16,Alcorn18}, which in turn can be compared with inferred baryonic masses to probe the amount of dark matter within galaxy effective radii. It will be possible to infer baryonic masses from the sum of gas masses estimated from dust-corrected star-formation rates, assuming the Schmidt-Kennicutt relation, and stellar masses estimated from WFIRST galaxy photometry. Such comparisons also place constraints on the allowed range of stellar initial mass functions \citep{Price16}.  Densely-sampled spectroscopy can also be used to study galaxy pairs and the evolution of the merger fraction and merger rate.  

Absorption line velocity dispersions, combined with galaxy sizes measured from WFIRST imaging, which enable the measurement of scaling relations and dynamical masses of spheroidal, quiescent galaxies. \citet{Belli17} and \citet{VanDokkum09} have used such measurements to demonstrate the evolution towards higher stellar velocity dispersion at fixed mass at the highest redshifts where such measurements have been performed to date (i.e., $z\sim 2$).

Small velocity shifts between ISM absorption lines (e.g., MgII 2800\AA, Na I 5890,5896\AA) and galaxy systemic redshifts (e.g., from H$\alpha$ or [OIII] emission) can be used to trace gas flows around star-forming galaxies.  Most kinematic evidence of this type suggests outflows \citep[e.g.,][]{Shapley03,Steidel10,Weiner09,Chen10}, but there are rare instances of kinematic evidence of infalling gas \citep[][]{Rubin12,Martin12}.

\section{Reionization and Cosmic Structure}
\label{sec:reionization}
\vspace{-0.5ex}

Current observations indicate that the intergalactic medium (IGM) completed its transition from neutral to ionized by $z=6.5$. This process is poorly understood: it is usually presumed that star-forming galaxies were responsible, but there is little evidence that sufficient ionizing radiation escapes from early galaxies to accomplish this.  Reionization may have been highly inhomogeneous as well, with expanding bubbles driven by strongly clustered young galaxies that are highly biased tracers of dark matter structure. In coming decades, new radio facilities (LOFAR, HERA, 
SKA) will map (at least statistically) the distribution of neutral hydrogen in the epoch of reionization; Wide-field infrared spectroscopy will provide essential complementary information about the spatial distribution of the (potentially) ionizing galaxies themselves over the same sky areas and redshift ranges. This requires accurate spectroscopy deep enough to detect [OIII] 4959,5007\AA\ emission at $5 < z < 7$ over very wide sky areas to characterize the clustering of early galaxies.   [OIII] emission is a signature of the low-metallicity population that may be the main driver of IGM reionization. An accurate measurement of 3D clustering will strongly constrain theoretical models that can then be extrapolated to higher redshifts, earlier in the epoch of reionization. 

There is already evidence that Ly$\alpha$ may be inhomogeneously suppressed by the neutral IGM (e.g., \cite{Tilvi14}), and that its escape may correlate with galaxy overdensities that can more effectively ionize large IGM volumes \citep{Castellano16}.  Ground-based optical observations with wide-field, high-multiplex spectrometers can measure Ly$\alpha$ out to $z \approx 7$, while higher redshifts require infrared spectroscopy.  Target galaxies may be selected from deep WFIRST Guest Observer science programs over many square degrees. With or without Ly$\alpha$, 3D correlation of large spectroscopic samples (e.g., from [OIII]) with 21-cm surveys of the neutral IGM in the same volumes will offer invaluable constraints on the reionization process. 
% (see Astro2020 white paper from Furlanetto et al.\ 2019).

\section{Required Measurements}
\vspace{-0.5ex}

While wide-field, high-multiplex fiber spectroscopy from ground-based telescopes (e.g., DESI, PSI, MOONS, MSE) will measure large galaxy samples at high redshifts, they are limited to optical or short-wavelength near-infrared observations.  Thermal background for infrared fiber spectroscopy, plus wavelength restrictions imposed by atmospheric absorption and emission will limit the redshift ranges over which galaxies can be observed, especially for  detection of multiple spectral features that can enable physical diagnostics of ISM conditions or stellar population properties.   Space-based infrared spectroscopy at wavelengths similar to those observed by JWST NIRSpec, with a dark orbital sky free of atmospheric limitations, would be ideal to measure galaxy emission lines and stellar continuum out to very high redshifts.

An example is ATLAS Probe (Wang et al.\ 2019), conceived as the spectroscopic follow-up to WFIRST. ATLAS Probe would use a 1.5m aperture telescope with an 0.4 deg$^2$ field of view and  Digital Micro-mirror Devices (DMDs) as slit selectors for $R = 1000$ spectroscopy at 1-4${\mu}$m with a target multiplex of $\sim$6000 per observation. ATLAS is designed to fit within the NASA probe-class space mission cost envelope. A nominal ATLAS mission would execute a set of three nested galaxy redshift surveys (wide: 2000 deg$^2$, medium: 100 deg$^2$, and deep: 1 deg$^2$), observing 200 million galaxies with $0.5<z<7+$ imaged by the WFIRST High Latitude Survey.

Understanding galaxy evolution in the context of large-scale structure is of critical importance in our quest to discover how the Universe works. The required observational facility can be established as part of the probe portfolio by NASA within the next decade.

\pagebreak


\begin{thebibliography}{}

\bibitem[Alcorn et al.(2018)]{Alcorn18}
Alcorn, L.~Y., Tran, K.-V., Glazebrook, K., et al.,
2018, ApJ, 858, 47
% "ZFIRE: 3D Modeling of Rotation, Dispersion, and Angular Momentum of Star-forming Galaxies at z ˜ 2."

\bibitem[Baldwin, Phillips, \& Terlevich(1981)]{Baldwin81}
Baldwin, J.A., Phillips, M.M., \& Terlevich, R. 1981, PASP, 93, 5

\bibitem[Behroozi et al.(2013)]{Behroozi13}
Behroozi, P., et al., 2013, ApJ, 770, 57

\bibitem[Behroozi et al.(2018)]{Behroozi18}
Behroozi, P., et al., 2018, arXiv e-prints, arxiv:1806.07893

\bibitem[Belli et al.(2017)]{Belli17}
Belli, S., et al., 2017, ApJ, 834, 18
%MOSFIRE Spectroscopy of Quiescent Galaxies at 1.5 < z < 2.5. I. Evolution of Structural and Dynamical Properties


\bibitem[Castellano et al.(2016)]{Castellano16}
Castellano, M., et al. 2016, ApJL, 818, L3


\bibitem[Chen et al.(2010)]{Chen10} Chen, Y.-M., Tremonti, C.~A., Heckman, T.~M., et al., 2010, AJ, 140, 445
% Absorption-Line Probes of the Prevalence and Properties of Outflows in Present-Day Star-forming Galaxies


\bibitem[Coil et al.(2011)]{Coil11}
Coil, A., et al., 2011, ApJ, 741, 8

\bibitem[Gobat et al.(2013)]{Gobat13} 
Gobat, R., Strazzullo, V., Daddi, E., et al. 2013, ApJ, 776, 9 

\bibitem[Guzzo et al.(2014)]{Guzzo14}
Guzzo, L., et al.\ 2014, A\&A, A108

\bibitem[Hearin et al.(2016)]{Hearin16}
Hearin et al. 2016, MNRAS, 461, 2135

\bibitem[Hudson et al.(2015)]{Hudson15} Hudson, M.~J., Gillis, B.~R., Coupon, J., et al.\ 2015, MNRAS, 447, 298 


\bibitem[Izquierdo et al.(2017)]{Izquierdo17}
Izquierdo-Villalba, D.; Orsi, A. A.; Bonoli, S.; Lacey, C. G.; Baugh, C. M.; Griffin, A. J., 2017,
 arXiv:1712.07129
%The environment of radio galaxies: A signature of AGN feedback at high redshifts
  
\bibitem[Kravtsov et al.(2013)]{Kravtsov13} Kravtsov et al. 2013, ApJ, 764, 31


\bibitem[Lee et al.(2017)]{Lee17} Lee, C.~T., Primack, J.~R., Behroozi, P., et al.\ 2017, MNRAS, 466, 3834 


\bibitem[Lilly et al.(2009)]{Lilly09}
Lilly, S.\ J., et al.\ 2009, ApJS, 182, 218


\bibitem[Martin et al.(2012)]{Martin12} Martin, C.~L., Shapley, A.~E., Coil, A.~L., et al., 2012, ApJ, 760, 127
% Demographics and Physical Properties of Gas Outflows/Inflows at 0.4<z<1.4


\bibitem[More et al.(2016)]{More16} More, S., Miyatake, H., Takada, M., et al.\ 2016, ApJ, 825, 39 


\bibitem[\protect\citeauthoryear{{Moster}, {Naab}  \& {White}}{{Moster}
  et~al.}{2013}]{Moster13}
{Moster} B.~P.,  {Naab} T.,   {White} S.~D.~M.,  2013, MNRAS{}
428, 3121


\bibitem[Orsi et al.(2016)]{Orsi16}
Orsi, {\'A}; Fanidakis, N.; Lacey, C. G.; Baugh, C. M., 2016, MNRAS, 456, 3827
%The environments of high-redshift radio galaxies and quasars: probes of protoclusters


\bibitem[Price et al.(2016)]{Price16}
Price, S.~H., Kriek, M., Shapley, A.~E. et al., 2016, 
ApJ, 819, 80
% The MOSDEF Survey: Dynamical and Baryonic Masses and Kinematic Structures of Star-forming Galaxies at $1.4 \leq z \leq 2.6$


\bibitem[Rubin et al.(2012)]{Rubin12} Rubin, K.~H.~R.,
Prochaska, J.~X., Koo, D.~C. et al., 2012, ApJL, 747, L26
% The Direct Detection of Cool, Metal-Enriched Gas Accretion onto Galaxies at z~0.5


\bibitem[Shapley et al.(2003)]{Shapley03}
Shapley, A.~E., Steidel, C.~C., Pettini, M., et al., 2003, ApJ, 588, 65
% Rest-frame Ultraviolet Spectra of z~3 Lyman Break Galaxies

    
\bibitem[Steidel et al.(2010)]{Steidel10}
Steidel, C.~C. et al., 2010, ApJ, 717, 289
%The Structure and Kinematics of the Circumgalactic Medium From Far-Ultraviolet Spectra of z~2-3 Galaxies


\bibitem[Tilvi et al.(2014)]{Tilvi14}
Tilvi, V., et al. 2014, ApJ, 794, 5


\bibitem[Valentino et al.(2015)]{Valentino15} 
Valentino, F., Daddi, E., Strazzullo, V., et al. 2015, ApJ, 801, 132 

\bibitem[Valentino et al.(2016)]{Valentino16} 
Valentino, F., Daddi, E., Finoguenov, A., et al. 2016, ApJ, 829, 53 


\bibitem[Van Dokkum et al.(2009)]{VanDokkum09}
Van Dokkum, P.~G., Kriek, M., Franx, M., 2009,
Nature, 460, 717
% A high stellar velocity dispersion for a compact massive galaxy at redshift z = 2.186

\bibitem[Vavrek et al.(2016)]{Vavrek16}
Vavrek, R.D., et al. (2016), Proc. of SPIE, Vol. 9911, 991105
%Mission-level performance verification approach for the Euclid space mission, 


\bibitem[Wang et al.(2016)]{WangT16} 
Wang, T., Elbaz, D., Daddi, E., et al. 2016, ApJ, 828, 56 

\bibitem[Wang et al.(2019)]{Wang19}
Wang, Y., et al.\ 2019, PASA, in press, arXiv:1802.01539v3

\bibitem[Weiner et al.(2009)]{Weiner09}
Weiner, B.~J., Coil, A.~L., Prochaska, J.~X., et al., 2009, ApJ, 692, 187
% Ubiquitous Outflows in DEEP2 Spectra of Star-forming Galaxies at z=1.4


\end{thebibliography}
\end{document}